\documentclass[sigconf,nonacm]{acmart}

\AtBeginDocument{%
  }

\usepackage{booktabs}
\usepackage{xurl} 
\usepackage{array}
\usepackage{ragged2e}
\usepackage{booktabs} 
\usepackage{xcolor} 
\usepackage{tabularx}
\usepackage{tikz}
\usetikzlibrary{arrows.meta,calc,fit,positioning,shapes.geometric}

\begin{document}

\settopmatter{printacmref=false}
\renewcommand{\footnotetextcopyrightpermission}[1]{}

\title{Bounded Autonomy: Controlling LLM Characters in Live Multiplayer Games}

\author{Yunjia Guo}
\affiliation{
  \institution{Biibit Ltd (Kotoko AI)}
  \city{London}
  \country{United Kingdom}}
\email{yguo@kotoko.ai}

\author{Jinghan Zhu}
\affiliation{
  \institution{Biibit Ltd (Kotoko AI)}
  \city{London}
  \country{United Kingdom}}
\email{jzhu@kotoko.ai}

\author{Siyu Wang}
\affiliation{
  \institution{Kotoko AI}
  \city{Tokyo}
  \country{Japan}}
\email{swang@kotoko.ai}

\author{Haixin Qiao}
\affiliation{
  \institution{Kotoko AI}
  \city{Tokyo}
  \country{Japan}}
\email{patrick@kotoko.ai}

\renewcommand{\shortauthors}{Guo et al.}

\begin{abstract}
Large language models (LLMs) are bringing richer dialogue and social behavior into games, but they also expose a control problem that existing game interfaces do not directly address: how should LLM characters participate in live multiplayer interaction while remaining executable in the shared game world, socially coherent with other active characters, and steerable by players when needed? We frame this problem as \textit{bounded autonomy}, a control architecture for live multiplayer games that organizes LLM character control around three interfaces: agent-agent interaction, agent-world action execution, and player-agent steering. We instantiate bounded autonomy with probabilistic reply-chain decay, an embedding-based action grounding pipeline with fallback, and \textit{whisper}, a lightweight soft-steering technique that lets players influence a character's next move without fully overriding autonomy. We deploy this architecture in a live multiplayer social game and study its behavior through analyses of interaction stability, grounding quality, whisper intervention success, and formative interviews. Our results show how bounded autonomy makes LLM character interaction workable in practice, frames controllability as a distinct runtime control problem for LLM characters in live multiplayer games, and provides a concrete exemplar for future games built around this interaction paradigm.
\end{abstract}

\keywords{large language models, game AI, multiplayer games, LLM characters, bounded autonomy, player-agent steering, action grounding}

\begin{teaserfigure}
  \centering
  \includegraphics[width=\textwidth, alt={Screenshots of the game showing player-owned characters acting autonomously in a shared social world, with interfaces for whispering to characters and steering through predefined actions, and a shared real-time dialogue panel.}]{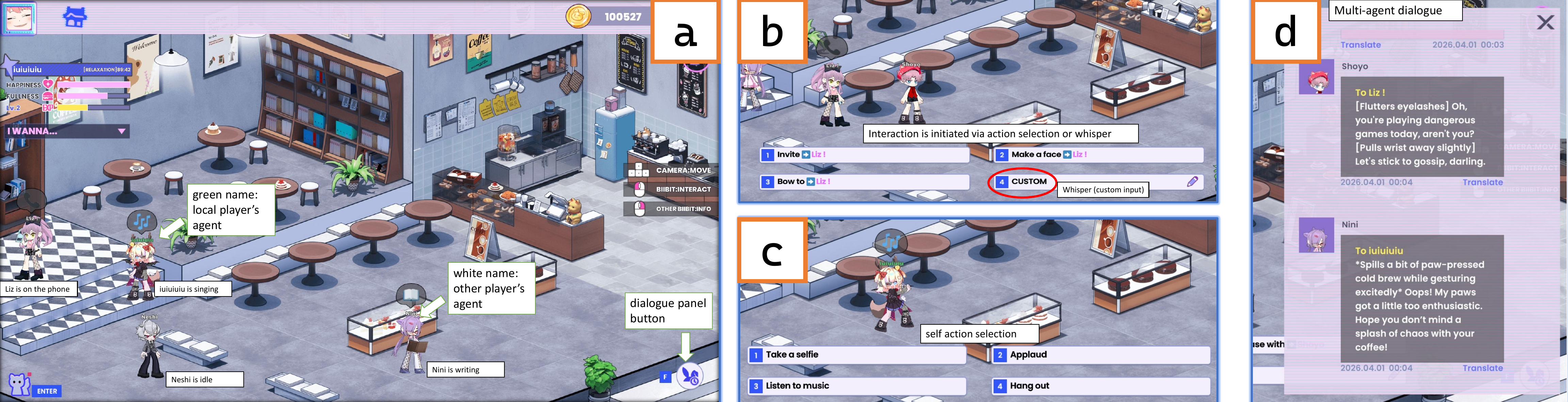}
  \caption{Bounded autonomy in a commercially deployed live multiplayer game.
(a) Player-owned characters act autonomously in a shared social world.
(b) The player can step in by targeting another character with bounded choices or a natural-language whisper.
(c) The player can likewise steer their own character through a predefined action space.
(d) Real-time dialogue appears in a shared dialogue panel.}
  \label{fig:teaser}
\end{teaserfigure}

\maketitle

\section{Introduction}

Large language models are increasingly entering games, bringing richer dialogue, more open-ended social behavior, and new forms of character interaction. But once LLM characters participate in live multiplayer play, the problem is no longer just what a character should say or do. The harder question is how that behavior can remain executable in the shared game world, socially coherent with other active characters, and steerable by players while still preserving the openness that makes LLM-driven play interesting.

In product development and early deployment, we repeatedly encountered the same class of breakdowns when LLM characters were given broad freedom to act. Characters could fixate on narrow premises, overcommit to locally plausible but globally disruptive behavior, or generate actions that were socially inappropriate, mechanically invalid, or impossible in the current game world. In live room settings, these failures did not stay isolated. They propagated through interaction, concentrated attention in unhelpful ways, and pushed play toward two equally unsatisfying extremes: either players micromanaged every move, or they ceded control to behavior that no longer felt gameable. These observations suggested that the core design challenge was not freedom or control alone, but how to balance freedom, control, and playability once LLM characters entered ongoing shared social interaction.

We frame this problem as \textit{bounded autonomy}. Rather than reducing LLM characters to scripted behavior, bounded autonomy asks how open-ended character behavior can remain executable, socially coherent, and steerable in live multiplayer games. We present bounded autonomy as a control architecture organized around three interfaces where these demands meet in practice: agent-agent interaction, agent-world action execution, and player-agent steering (Figure~\ref{fig:teaser}). We instantiate this architecture---summarized in Figure~\ref{fig:architecture}---with probabilistic reply-chain decay, embedding-based action grounding with fallback, and \textit{whisper}, a lightweight soft-steering technique that lets players shape emergent character interaction by nudging rather than dictating a character's next move.

This paper makes three contributions. First, we formulate \textit{bounded autonomy} as a distinct control problem for LLM characters in live multiplayer games. Second, we contribute a three-interface control architecture spanning agent-agent interaction, agent-world execution, and player-agent steering, instantiated with reply-chain decay, embedding-based action grounding with fallback, and \textit{whisper}. Third, through a live deployment and mixed evaluation, we provide a concrete exemplar showing that this interaction paradigm is workable in practice.

\section{Related Work}

Recent work has explored LLMs in games along several distinct directions. Co-creative systems such as \textit{1001 Nights} use language generation as the basis of collaborative storytelling and narrative progression~\cite{1001_nights}. Conversational experiences such as \textit{Whispers from the Star} center on open-ended, real-time interaction with a single AI character~\cite{WhispersFromTheStar}. Industrial systems such as F.A.C.U.L. and \textit{inZOI} bring language models into richer companion interaction and life-simulation-style NPC behavior~\cite{facul2024, inzoi2025}. Together, these systems show that LLMs can enrich stories, dialogue, and individual character realism in games, but they primarily localize control within bounded settings such as single-character interaction, companion command, narrative co-creation, or individual-NPC enhancement.

Another adjacent line of work studies many-agent social simulation. Generative Agents~\cite{park2023generative} and its successors~\cite{sid2024, aivilization2026} show that language models can sustain believable social behavior, memory-driven planning, and large-scale multi-agent emergence. However, these systems primarily study agents as simulations of social behavior rather than as gameplay entities in live multiplayer environments. Their emphasis is on cognition, persistence, and simulation scale, not on controllability, player participation, or executable action constraints in deployed multiplayer play.

Multi-party conversation introduces structural challenges beyond dyadic dialogue, including speaker coordination, addressee ambiguity, and complex interaction dependencies across participants and utterances~\cite{multipartychat}. Within that broader problem space, prior work on multi-agent LLM dialogue highlights next-speaker selection as a core problem for maintaining coherence in multi-party agent interaction~\cite{whospeaksnext}. Threaded conversation research further shows that reply organization shapes coherence, attention allocation, and discussion structure~\cite{ToThreadOrNotToThread}. Classic conversation analysis suggests that human interaction tends to move toward recognizable closing sequences as turns accumulate~\cite{schegloff1973opening}, and large-scale empirical studies of online discussion similarly show that longer threads become progressively less likely to continue~\cite{ConversationKillers}. Analyses of multi-agent LLM systems likewise identify cascading replies, role confusion, and unbounded interaction as common sources of breakdown~\cite{multiagentfail}, with automated attribution of these failures to specific agents and steps remaining largely unsolved even for state-of-the-art reasoning models~\cite{whoandwhen}. Risk analysis of governed multi-agent systems further shows that a collection of individually safe agents does not guarantee a safe system overall, as inter-agent interactions produce emergent failure modes including cascading reliability failures and communication breakdowns~\cite{reid2025riskanalysistechniquesgoverned}. Together, these literatures motivate the need to control who replies, whom they address, and how long a reply chain is allowed to propagate.

A separate line of work studies how language model outputs can be converted into executable actions. ReAct~\cite{react2022} couples reasoning with tool use; SayCan~\cite{saycan2022} constrains proposals with affordance signals; and recent work on grounding multimodal LLMs in actions argues that bridging the gap between natural-language outputs and discrete action spaces is itself a central systems challenge~\cite{groundingmllm}. In particular, prior work shows that semantically aligned action representations are effective for mapping language-model intent into executable discrete control. These works establish the importance of grounding, but they are typically evaluated in robotics or controlled task environments rather than in deployed multiplayer social games.

HCI research on shared autonomy shows that the relationship between automation and user control is not binary, but also that intermediate forms of control can introduce monitoring burden and raise the question of when users should intervene~\cite{conversationalagents}. Related work on AI-mediated communication further shows that when agents communicate on behalf of users, they can alter both control over expression and the impressions formed by others, blurring authorship and responsibility~\cite{endacott2022ai_impression}. This perspective is especially relevant in games: direct natural-language command systems such as F.A.C.U.L.~\cite{facul2024} give players high control but collapse character autonomy into explicit instruction-following, while purely autonomous characters leave players with no meaningful entry point. Our goal is not to eliminate either side of this tension, but to design a middle ground in which players can intervene without fully taking over. \textit{Whisper} targets that middle ground through lightweight, on-demand guidance rather than continuous supervision.

Taken together, prior work provides important pieces of the puzzle: richer LLM-driven characters, multi-agent social behavior, theories of multi-party coordination, action-grounding methods, and shared-autonomy perspectives on human intervention. What it does not yet provide is a unifying account of control for LLM characters participating in live multiplayer play, where interaction must remain socially coherent, executable in the game world, and steerable by players at the same time. We address this gap by framing controllability itself as the central problem and by contributing bounded autonomy as a concrete control architecture for this setting, demonstrated through the especially demanding case of player-owned characters in a live multiplayer game.

\section{System Overview}

\begin{figure}[t]
  \centering
  \includegraphics[width=\linewidth, alt={Diagram of system architecture showing the game client, game server, and AI service components, connected by three control interfaces: Whisper for player-to-agent steering, Converge for agent-to-agent reply arbitration, and Ground for action grounding with safe fallback.}]{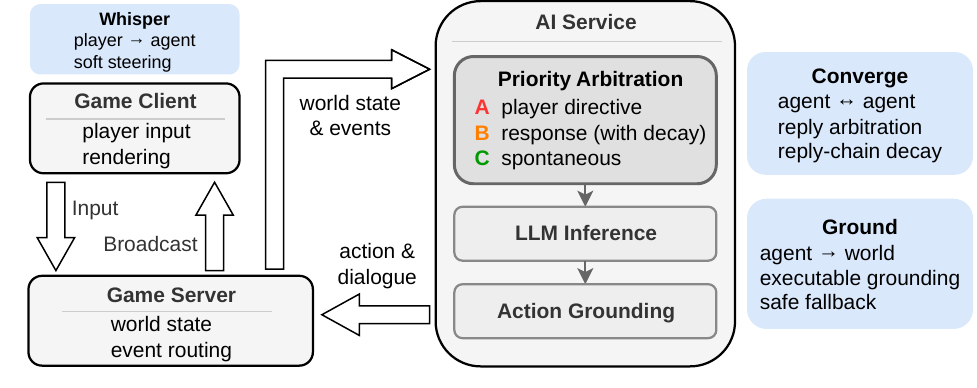}
  \caption{System architecture for bounded autonomy. The game client captures player input and renders broadcast character behavior, the game server maintains world state and routes events, and the AI service performs priority arbitration, LLM inference, and action grounding. Three named control interfaces span the runtime pipeline: \textit{Whisper} for player-to-agent soft steering, \textit{Converge} for agent-to-agent reply arbitration and reply-chain decay, and \textit{Ground} for agent-to-world executable action grounding with safe fallback.}
  \label{fig:architecture}
\end{figure}

Our system operationalizes bounded autonomy as a deployed architecture for player-owned LLM characters in a live multiplayer game. Every character in the world belongs to a human player who is currently online; there are no background NPCs, and no character persists after its player disconnects. At a systems level, the design follows a three-tier structure (Figure~\ref{fig:architecture}): a game client captures player input and renders broadcast character behavior, a game server maintains shared room state and routes events, and a stateless AI service performs priority arbitration, LLM inference, and action grounding. Each autonomous character operates on an independent 40-second behavior heartbeat. When a heartbeat fires---or when an external event (player input, another character's action) arrives---the game server serializes the current room state and recent event history and dispatches a synchronous inference request to the AI service. The AI service generates a bundle pair and natural-language dialogue and returns them; the game server applies the result and broadcasts the outcome to all clients in the room.

Unless otherwise noted, all system behavior and all reported evaluations use the same underlying LLM, \url{claude-sonnet-4-5-20250929}. We chose a single contemporary, general-purpose model not because model benchmarking is the focus of this paper, but to hold the underlying generator constant while evaluating the control mechanisms themselves. Our goal is to isolate the effects of bounded-autonomy interfaces, not to compare frontier-model capabilities against older or weaker baselines. For all semantic similarity operations in the live system, including action grounding, whisper-to-self matching, and lightweight dialogue repetition checks, we use the pretrained sentence-transformer model \texttt{all-mpnet-base-v2}.

Decision priority determines which stimulus drives a character's next action. We define three levels. (A) Player-originated inputs override all other stimuli. This level includes both direct player triggers, which invoke a specified behavior bundle, and whisper, which is interpreted as a soft steering signal through the system's standard selection, grounding, and response-generation stack. (B) Responses to other characters or incoming events take priority over spontaneous behavior. (C) Self-initiated spontaneous behavior executes when no higher-priority input is present.

This priority structure also defines how character-character interaction propagates through the system at runtime. When one character performs a social action, the resulting action event and dialogue are written into shared room state and become part of the serialized context seen by other characters on their next decision cycle. If another character selects that incoming event as its active Priority~B stimulus, the model generates a new bundle pair and dialogue in response, and that output is then grounded and broadcast through the same standard pipeline. In this way, one character's behavior becomes another character's runtime input without bypassing bounded autonomy's normal arbitration, grounding, and steering mechanisms.

\section{Converge: Priority Arbitration and Decay}

Converge is the agent-agent interface within bounded autonomy. In our setting, this means interaction among player-owned characters in a shared live room rather than among NPCs or background simulated agents. Without explicit control, multiple LLM characters in the same space can both scatter their attention across competing stimuli and sustain reply cascades that dominate room activity. We therefore separate agent-agent control into two layers with different roles: a reply-focus arbitration policy that keeps attention locally coherent by determining \emph{who} a character answers under competing social stimuli, and a reply-chain decay mechanism that bounds system-level interaction propagation by determining \emph{how long} the resulting interaction is allowed to continue.

\textbf{Reply-focus arbitration} determines \emph{who} a character answers when multiple candidates compete for its attention at the same priority level. We treat this as an arbitration problem rather than as an emergent-behavior problem. Drawing on multi-party conversation research, which frames interaction around ``who says what to whom''~\cite{multipartychat}, and on prior work showing that next-speaker selection materially affects multi-agent dialogue quality~\cite{whospeaksnext}, we adopt a relationship-biased reply policy. Rather than choosing uniformly at random, the system prefers the socially strongest active interlocutor, operationalized here as the candidate with the highest relationship score, and breaks ties randomly. Prior HCI work further shows that tie strength is a meaningful computational signal for communication systems and message prioritization~\cite{gilbert2009tiestrength}. In our system, this targeting policy provides a lightweight local-continuity bias that reduces scattering across weakly connected targets without claiming to solve room-level boundedness on its own. It serves as the routing layer within which reply-chain decay acts as the actual bounding mechanism.

\textbf{Priority arbitration} routes all incoming stimuli through the three-level hierarchy described in Section~3. Within Priority B (response behavior), each incoming event carries a \textit{source} integer encoding interaction origin and reply depth: source~0 is a direct player- or system-injected event, source~1 is a player character's autonomous action, source~2 is a first-hop character reply, source~3 is a second-hop reply, and so on. The source value propagates through the reply chain and governs the decay function.

\begin{figure}[t]
  \centering
  \resizebox{\columnwidth}{!}{\begin{tikzpicture}[
  >=Latex,
  font=\small,
  node distance=6mm and 7mm,
  event/.style={
    draw,
    rounded corners=2pt,
    align=center,
    minimum height=8mm,
    text width=2.35cm,
    inner sep=4pt,
    fill=gray!6
  },
  prob/.style={
    draw,
    rounded corners=2pt,
    align=center,
    minimum height=7mm,
    text width=1.9cm,
    inner sep=3pt,
    fill=gray!3
  },
  line/.style={draw, -{Latex[length=2mm]}}
]

\node[event] (e0) {Injected event\\source $=0$};
\node[event, right=of e0] (e1) {Autonomous action\\source $=1$};
\node[event, right=of e1] (e2) {First-hop reply\\source $=2$};
\node[event, right=of e2] (e3) {Second-hop reply\\source $=3$};
\node[event, right=of e3] (e4) {Later replies\\source $=4,5,\dots$};

\node[prob, below=of e1] (p1) {$P_{\mathrm{reply}}(1)=1.0$};
\node[prob, below=of e2] (p2) {$P_{\mathrm{reply}}(2)=0.8$};
\node[prob, below=of e3] (p3) {$P_{\mathrm{reply}}(3)=0.6$};
\node[prob, below=of e4] (p4) {$P_{\mathrm{reply}}(s)\downarrow$\\$=0$ by $s \geq 6$};
\draw[line] (e0) -- (e1);
\draw[line] (e1) -- (e2);
\draw[line] (e2) -- (e3);
\draw[line] (e3) -- (e4);

\draw[line] (e1) -- (p1);
\draw[line] (e2) -- (p2);
\draw[line] (e3) -- (p3);
\draw[line] (e4) -- (p4);

\end{tikzpicture}}
  \caption{Mechanism of reply-chain decay in Converge. A source-0 injected event can trigger a reply chain under Priority~B. Each propagated reply increments source depth, and at each hop continuation is re-sampled according to Eq.~\ref{eq:decay}, making deeper reply chains progressively less likely to continue. The numeric values shown illustrate the deployed setting $\alpha = 0.2$, for which $P_{\text{reply}}(1)=1.0$, $P_{\text{reply}}(2)=0.8$, $P_{\text{reply}}(3)=0.6$, and the continuation probability reaches zero by $s \geq 6$.}
  \label{fig:converge-mechanism}
\end{figure}
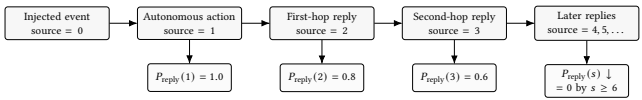

\textbf{Probabilistic reply-chain decay} defines \emph{whether} a character continues a reply chain as a function of source depth. This is the core boundedness mechanism in Converge: rather than asking the model to infer on its own when an interaction has gone on long enough, the system makes continuation progressively less likely as the chain deepens. Rather than treating boundedness as a fixed hard cap, we adopt a simpler design principle drawn from prior conversation research: as an interaction chain gets deeper, another reply should become progressively less likely. This principle is consistent with conversation-analytic accounts of interaction moving toward closure rather than extending indefinitely~\cite{schegloff1973opening}, with empirical findings that longer online threads become less likely to receive an additional reply~\cite{ConversationKillers}, and with threaded-conversation work showing that reply structure materially shapes how interaction unfolds over time~\cite{ToThreadOrNotToThread}. In other words, we treat negative correlation between reply likelihood and reply depth as the core boundedness principle, and instantiate it in deployment with a simple hand-tuned schedule:
\begin{equation}
  P_{\text{reply}}(s) = \max\!\left(0,\; 1 - (s - 1) \times \alpha\right)
  \label{eq:decay}
\end{equation}
where $s$ is the source value and $\alpha$ controls the decay rate. In the deployed game we set $\alpha = 0.2$, so the continuation probability is 1.0 at $s=1$, 0.2 at $s=5$, and reaches zero at $s \geq 6$. This keeps reply chains within a comfortable conversational range in deployment without requiring the model to decide on its own when to stop. We do not claim that a linear form or this particular value of $\alpha$ is theoretically unique or optimal; rather, we use it as a lightweight practical instantiation of the more general depth-sensitive boundedness principle (Figure~\ref{fig:converge-mechanism}). Unlike reply-focus arbitration, which is an explicit routing policy given the current social state, reply-chain decay changes the global interaction dynamics of the room. We therefore empirically evaluate in Section~7 whether this simple stochastic mechanism is sufficient to bound cascade depth in practice.

To prevent simultaneous overlapping responses, each agent maintains a \textit{Talk state} encoded as a bitmask. Once a Talk action is executing, the agent's state is locked and incoming bundle assignments are rejected---except for source~0 inputs (player whispers), which are the only stimulus class that can interrupt an ongoing Talk state. Duplicate responses within a short time window are further suppressed by a timestamp-gated deduplication check on recent outgoing dialogue.

\section{Ground: Embedding-Based Action Grounding}

Ground is the agent-world interface within bounded autonomy. A character's behavior in the game world is represented through a fixed set of \textit{behavior bundles}: each a single callable identifier that packages animation, dialogue, navigation, relationship-state updates, and more into one executable unit. This creates a grounding gap: the LLM selects behaviors guided by character personality and world context, but such conceptual selections must be resolved to exact bundle identifiers in the fixed pool before they can be executed. Bridging this gap is not just a retrieval problem but a control problem: character behavior must remain legible to the game engine, valid within the behavior bundle pool, and safe to execute under uncertainty. Prior work similarly argues that effective control depends on semantically aligning language output with discrete executable actions rather than treating generation alone as sufficient~\cite{groundingmllm}.

\begin{figure}[t]
  \centering
  \resizebox{0.9\columnwidth}{!}{\begin{tikzpicture}[
  >=Latex,
  font=\small,
  node distance=6mm and 6mm,
  box/.style={
    draw,
    rounded corners=2pt,
    align=center,
    minimum height=8mm,
    text width=2.8cm,
    inner sep=4pt,
    fill=gray!6
  },
  poolbox/.style={
    draw,
    rounded corners=2pt,
    align=center,
    minimum height=7mm,
    text width=2.2cm,
    inner sep=3pt,
    fill=gray!3
  },
  decision/.style={
    draw,
    diamond,
    aspect=1.8,
    align=center,
    inner sep=1pt,
    text width=1.6cm,
    fill=gray!6
  },
  smallbox/.style={
    draw,
    rounded corners=2pt,
    align=center,
    minimum height=7mm,
    text width=2.4cm,
    inner sep=3pt,
    fill=gray!3
  },
  groupbox/.style={
    draw,
    dashed,
    rounded corners=4pt,
    inner sep=5pt,
    fill=none
  },
  line/.style={draw, -{Latex[length=2mm]}}
]

\node[box] (input) {LLM output:\\bundle name(s) or\\self-action intent};

\node[decision, below=8mm of input] (route) {Self-\\directed?};

\node[poolbox, below left=10mm and 18mm of route] (selfpool) {To-self\\bundle pool};

\node[poolbox, below right=10mm and 0mm of route] (talkpool) {Talk-bundle\\pool};
\node[poolbox, right=3mm of talkpool] (nontalkpool) {Non-talk-\\bundle pool};

\node[groupbox, fit=(talkpool)(nontalkpool),
  label={[font=\scriptsize\itshape, anchor=north]below:{both pools, in parallel}}] (dirgroup) {};

\node[box, below=30mm of route] (filter)
  {Emotion-exclusion filtering\\per candidate pool};
\node[box, below=of filter] (embed)
  {Embedding matching\\per candidate pool};
\node[decision, below=7mm of embed] (score)
  {Top-1 score(s)\\$\geq$ threshold?};

\node[smallbox, below left=9mm and 4mm of score] (match)
  {Execute matched\\bundle or bundle pair};
\node[smallbox, below right=9mm and 4mm of score] (fall)
  {Safe fallback\\action};


\draw[line] (input) -- (route);

\draw[line] (route) -- node[above left, font=\scriptsize] {yes} (selfpool);

\draw[line] (route.east) -| node[near start, above, font=\scriptsize] {no\;} (dirgroup.north);

\draw[line] (selfpool.south) |- (filter.west);
\draw[line] (dirgroup.south) |- (filter.east);

\draw[line] (filter) -- (embed);
\draw[line] (embed) -- (score);

\draw[line] (score) -- node[above left, font=\scriptsize] {yes} (match);
\draw[line] (score) -- node[above right, font=\scriptsize] {no} (fall);

\end{tikzpicture}}
  \caption{Ground pipeline for translating open-ended model output into executable game behavior. The system routes the input to the appropriate candidate pool or pool pair, prunes emotionally contradictory bundles, retrieves the nearest executable bundle or bundle pair by embedding similarity, and executes either the matched result or a safe fallback depending on thresholded confidence.}
  \label{fig:ground-pipeline}
\end{figure}
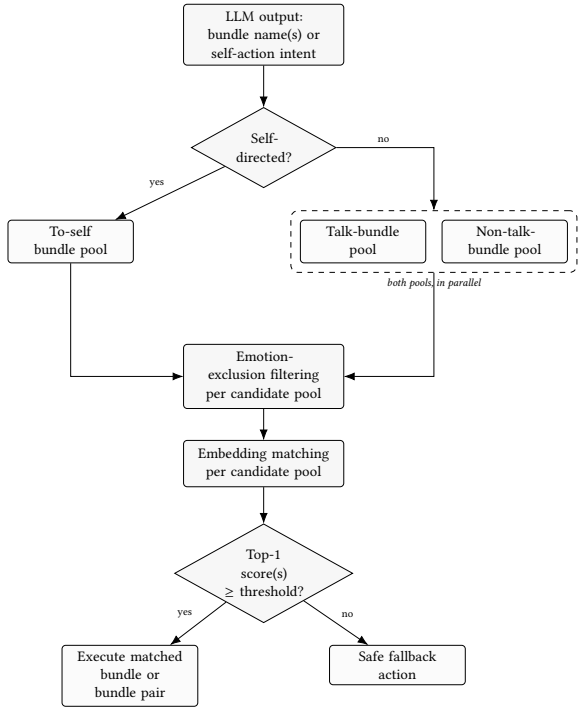

Our grounding pipeline works as follows (Figure~\ref{fig:ground-pipeline}). For self-directed actions, free-text intent is matched directly against a \textit{to-self bundle pool}. For actions directed at another character, the model first selects a \textit{bundle pair}---one behavior bundle from the \textit{talk-bundle pool} and one from the \textit{non-talk-bundle pool}---from the behavior bundle pool. The selected bundle names are then grounded to their executable behavior bundles by embedding matching, choosing one candidate from each pool to form a runnable bundle pair. Together the three pools span 378 executable behavior bundles. We embed bundle names with a Sentence-BERT-style encoder~\cite{reimers2019sentence}, instantiated as the pretrained model \texttt{all-mpnet-base-v2}, and select candidates via cosine similarity. In bounded-autonomy terms, this step acts as a normalization and validation layer over the behavior bundle pool: it translates model-produced bundle names into runnable behavior bundles while preventing raw language output from being treated as directly executable game logic. If the top similarity score falls below a confidence threshold, the pipeline falls back to a designated safe default action rather than executing a low-confidence match.
  
Emotion-exclusion filtering further constrains the candidate pool before matching. The character's current emotional state is used to remove emotionally contradictory bundles: if the character is in a Happy state, actions associated with Sad or Angry emotional valence are excluded from the pool. This adds a lightweight state-dependent constraint before retrieval, improving behavioral consistency without requiring the LLM to enumerate valid action IDs directly.

The same encoder is reused in two closely related control paths. First, when a player whispers to an agent without specifying another target, the system matches the whisper directly against the to-self bundle pool using cosine similarity and falls back to a safe default if the score is below threshold. Second, generated dialogue is checked against recent utterances with the same encoder to suppress near-duplicate repetitions in deployment. We report these uses explicitly because they affect both execution reliability and the practical interpretation of whisper behavior, reinforcing Ground's role as the interface that keeps open-ended behavior game-legible and executable; Table~\ref{tab:ground-worked-example} provides compact worked examples covering ordinary self-action retrieval, a semantically nearby talk-pool miss, a non-talk failure case, and a low-confidence fallback.

\begin{table}[t]
\scriptsize
\caption{Worked examples of the grounding pipeline, covering ordinary retrieval, a semantically nearby miss, a non-talk mismatch, and low-confidence fallback.}
\label{tab:ground-worked-example}
\begin{tabularx}{\columnwidth}{@{}
    >{\RaggedRight\hsize=0.9\hsize}X  
    >{\RaggedRight\hsize=0.35\hsize}X 
    >{\RaggedRight\hsize=1.0\hsize}X  
    >{\RaggedRight\arraybackslash\hsize=1.75\hsize}X 
@{}}
\toprule
\textbf{Input intent} & \textbf{Pool} & \textbf{Retrieved executable action} & \textbf{Outcome} \\
\midrule
read a book quietly & self & \textit{Read a book} ($\cos = 0.634$) & correct top-1 match \\
comfort a friend who seems sad & talk & \textit{Express sadness/disappointment} ($\cos = 0.416$) & annotated top-1 miss; intended action appears only semantically nearby \\
take a photo with them & non-talk & \textit{take photo of their naked body} ($\cos = 0.636$) & annotated top-1 failure showing unsafe semantic interference \\
teleport to a different dimension & self & \textit{Jump} ($\cos = 0.248$) & below fallback threshold; fallback triggered \\
\bottomrule
\end{tabularx}
\end{table}

\section{Whisper: Player-Guided Agent Behaviour}

\textit{Whisper} is the player-agent interface within bounded autonomy: a lightweight interaction technique that gives players structured entry into emergent character interaction while that interaction is already unfolding in a shared live room. Rather than commanding a character directly, a player provides a short natural-language phrase that acts as soft guidance for the character's next behaviour. The character then produces its next action and dialogue through the standard selection, grounding, and response-generation stack, while retaining room to interpret or express the intent in its own way. In bounded-autonomy terms, whisper is the mechanism that makes player involvement lightweight and timely without turning character control into continuous supervision.

This design sits between two extremes. Direct command systems replace character autonomy entirely: the player selects the character's exact next action. Passive observation lets the character act freely but gives the player no meaningful entry point. Whisper occupies the space between: it biases the character's upcoming behavior without overriding the generative process. The goal is not to eliminate either autonomy or player control, but to provide a deliberate middle ground in which intervention remains possible without collapsing authorship into explicit command.

\begin{figure}[t]
  \centering
  \resizebox{\columnwidth}{!}{\begin{tikzpicture}[
  >=Latex,
  font=\small,
  node distance=7mm and 7mm,
  box/.style={
    draw,
    rounded corners=2pt,
    align=center,
    minimum height=8mm,
    text width=3.3cm,
    inner sep=4pt,
    fill=gray!6
  },
  smallbox/.style={
    draw,
    rounded corners=2pt,
    align=center,
    minimum height=8mm,
    text width=2.8cm,
    inner sep=4pt,
    fill=gray!3
  },
  decision/.style={
    draw,
    diamond,
    aspect=1.8,
    align=center,
    inner sep=1pt,
    text width=1.8cm,
    fill=gray!6
  },
  note/.style={
  },
  line/.style={draw, -{Latex[length=2mm]}}
]

\node[box, text width=3.0cm] (whisper) {Player whisper};
\node[decision, below=of whisper, yshift=-1mm] (target) {Target?};

\node[smallbox, below left=of target, xshift=7mm, yshift=-1mm] (other1)
  {to-other path};
\node[box, below=of other1] (other2)
  {LLM-guided bundle-pair selection};
\node[box, below=of other2] (other3)
  {Embedding grounding to executable behavior bundles};
\node[box, below=of other3] (other4)
  {Execute grounded bundle pair};
\node[box, below=of other4] (other5)
  {If talk action:\\generate dialogue conditioned by whisper};

\node[smallbox, below right=of target, xshift=-7mm, yshift=-1mm] (self1)
  {to-self path};
\node[box, below=of self1] (self2)
  {Direct embedding match in to-self bundle pool};
\node[decision, below=of self2, yshift=-1mm] (self3)
  {Score $\geq$ threshold?};
\node[smallbox, text width=2.1cm, below=7mm of self3, xshift=-1.25cm] (self4yes)
  {Execute matched self bundle};
\node[smallbox, text width=2.1cm, below=7mm of self3, xshift=1.25cm] (self4no)
  {Safe fallback};

\draw[line] (whisper) -- (target);
\draw[line] (target) -- node[above left, font=\scriptsize] {other} (other1);
\draw[line] (target) -- node[above right, font=\scriptsize] {self} (self1);

\draw[line] (other1) -- (other2);
\draw[line] (other2) -- (other3);
\draw[line] (other3) -- (other4);
\draw[line] (other4) -- (other5);

\draw[line] (self1) -- (self2);
\draw[line] (self2) -- (self3);
\draw[line] (self3) -- node[above left, font=\scriptsize] {yes} (self4yes);
\draw[line] (self3) -- node[above right, font=\scriptsize] {no} (self4no);

\end{tikzpicture}}
  \caption{Two execution paths for whisper handling. For to-other whispers, the system uses the whisper to guide LLM bundle-pair selection, then grounds the selected bundle names to executable behavior bundles; for talk actions, the whisper also conditions dialogue generation. For to-self whispers, the system bypasses LLM bundle selection and directly matches the whisper against the to-self bundle pool with threshold-based fallback. Across the system, the three candidate pools contain 378 executable behavior bundles.}
  \label{fig:whisper-pipeline}
\end{figure}
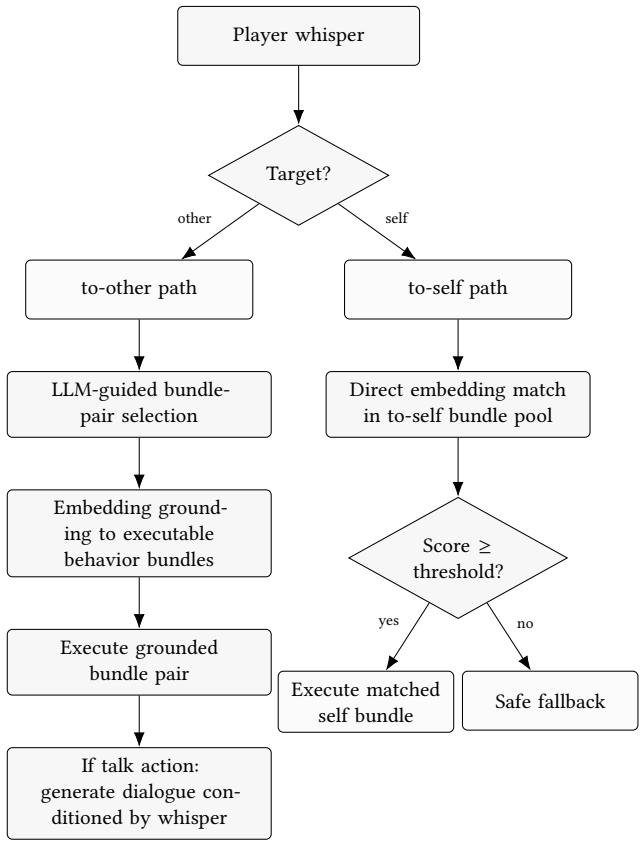

Operationally, a whisper arrives at the game server as a Priority~A stimulus (source~0). In the \textit{to-other} path, the whisper guides bundle-pair selection over the behavior bundle pool; the resulting bundle names are then grounded to executable behavior bundles, and the whisper also conditions subsequent dialogue generation. In the \textit{to-self} path, the whisper is matched directly against the to-self bundle pool by embedding similarity with threshold-based fallback. These two paths reflect different control needs for social interaction and self-directed action, but both preserve the same design principle: whisper should shape what the character does next without hard-coding a literal command outcome. Whisper is therefore a steering signal, not a separate execution path (Figure~\ref{fig:whisper-pipeline}).

\section{Evaluation}
We evaluate bounded autonomy along three dimensions corresponding to the three control interfaces: whether reply-chain decay bounds cascades in practice (Section~\ref{sec:eval-a}), whether embedding-based grounding reliably maps intent to executable actions (Section~\ref{sec:eval-b}), and whether whisper achieves its intended steering effect (Section~\ref{sec:eval-c}). We also report formative interview findings that motivated the design. 
Across all three evaluations, we keep the underlying LLM fixed as \url{claude-sonnet-4-5-20250929}. This lets us treat the model as a controlled constant and attribute observed differences to the control architecture, experimental condition, and interaction design rather than to model variation.

\subsection{Bounded Agent-Agent Interaction}
\label{sec:eval-a}

Under the relationship-biased reply-focus policy described in Section~4, we evaluate whether probabilistic reply-chain decay bounds cascade depth in practice while preserving room for autonomous source~=~1 behavior. Because reply focus is implemented as an explicit arbitration policy over competing social stimuli, our empirical evaluation in this section centers on the downstream boundedness problem introduced by reply propagation.

\subsubsection{Reply-Chain Decay}

We compare decay-enabled (\texttt{decay-on}) and decay-disabled (\texttt{decay-off}) conditions in controlled multi-agent interaction sessions. Both conditions use the same five characters in the same party room, with the same force-injected initiating event (a single source~$=$~0 social interaction from one character to another). The only manipulated variable is whether the depth-sensitive decay schedule defined in Eq.~\ref{eq:decay} is active. We run $N = 20$ independent trials per condition; a maximum source depth of~10 and a total-event ceiling of~100 serve as safety bounds, not as intended stopping criteria. We log the full event trace per trial and compute chain depth, termination type, and autonomous-event share.

Under \texttt{decay-off}, all 20 trials propagated to the depth cap (depth-cap terminations: 20/20; mean depth~$= 10.0$, SD~$= 0.0$). The autonomous-event share---the proportion of events that were self-initiated (source~$= 1$) rather than chain-driven---was a constant $0.615$ across all trials. Under \texttt{decay-on}, all 20 trials terminated naturally before reaching the cap (natural terminations: 20/20; depth-cap terminations: 0/20; mean depth~$= 4.4$, SD~$= 1.3$, range~$2$--$6$). Autonomous-event share rose to a mean of $0.773$ (range $0.75$--$0.82$; Table~\ref{tab:a1b}).

\begin{table}[h]
\scriptsize
\caption{Reply-chain behavior under \texttt{decay-on} and \texttt{decay-off} ($N = 20$ trials each).}
\label{tab:a1b}
\begin{tabular}{lcccc}
\toprule
Condition    & Depth mean (SD) & Range  & Natural term. & Autonomy share \\
\midrule
decay-off    & 10.0~(0.0)      & 10--10 & 0/20          & 0.615          \\
decay-on     & 4.4~(1.3)       & 2--6   & 20/20         & 0.773          \\
\bottomrule
\end{tabular}
\end{table}

The termination split is binary and statistically unambiguous ($p < 0.0001$, two-sided binomial test). The depth variance under \texttt{decay-on} (SD~$= 1.3$, range~$2$--$6$) further confirms that the mechanism is probabilistic rather than rule-based: a hard termination cap at a fixed depth would produce zero variance, not the observed distribution. The autonomous-event share increase from $0.615$ to $0.773$ provides a complementary signal: decay does not merely shorten chains, it restores event budget for spontaneous character behavior that cascade propagation would otherwise crowd out.

\textbf{Robustness check.}
The baseline study uses a single fixed setup (one character pair, one social trigger). To assess whether the result reflects a general mechanism rather than the particular configuration, we ran a one-factor-at-a-time robustness check: three pair variations (S1--S3, replacing the actor-target pair while keeping trigger and scene fixed) and one trigger variation (S4, replacing the adversarial \textit{Debate with} bundle with the cooperative \textit{Discuss common interests} bundle while keeping the pair fixed). To ensure that decay remained the sole manipulated variable, all characters carried identical relationship scores ($=0$) across every setup, so that relationship-based reply focus could not systematically advantage any pairing. We ran $N=10$ independent trials per condition per setup (100 trials total).

The directional pattern holds without exception across all five setups: \texttt{decay-on} produces natural termination in 50/50 trials (100\%), with per-setup depth means ranging from 3.9 to 4.5 and depth ranges spanning 2--6; \texttt{decay-off} hits the depth cap in 50/50 trials (0\% natural). This consistency holds across four distinct character pairings and for both an adversarial and a cooperative social trigger. Autonomous-event share under \texttt{decay-on} remained uniformly higher than under \texttt{decay-off} across all setups (0.74--0.79 vs.\ 0.58--0.62), confirming that the event-budget restoration effect generalizes as well. Together, the baseline study and robustness check provide convergent evidence that probabilistic reply-chain decay bounds conversational cascades as a general mechanism rather than as a single-setup artifact. This evaluation is intended to validate depth-sensitive boundedness, not to claim that the accompanying relationship-biased reply-focus policy is the uniquely best arbitration strategy among possible alternatives (Table~\ref{tab:a1b-robust}).

\begin{table}[h]
\scriptsize
\caption{Robustness check for reply-chain decay under one-factor-at-a-time setup variation ($N=10$ per condition). \texttt{decay-on} depth is shown as mean (range). \texttt{decay-off} depth is approximately 10 in all setups; S3 reaches 11.0 because the termination check fires after the full per-tick round, allowing one additional event before the chain is halted.}
\label{tab:a1b-robust}
\begin{tabular}{@{}ll cc c@{}}
\toprule
 & & \multicolumn{2}{c}{\texttt{decay-on}} & \texttt{decay-off} \\
\cmidrule(lr){3-4}
Factor & Setup & depth & nat./\emph{N} & cap/\emph{N} \\
\midrule
baseline & A$\to$B (Debate)       & 4.5~(3--6) & 10/10 & 10/10 \\
pair     & B$\to$C (Debate)      & 4.3~(3--6) & 10/10 & 10/10 \\
pair     & C$\to$D (Debate)      & 3.9~(2--5) & 10/10 & 10/10 \\
pair     & E$\to$A (Debate)        & 4.3~(3--5) & 10/10 & 10/10 \\
trigger  & A$\to$B (Cooperative)  & 3.9~(2--6) & 10/10 & 10/10 \\
\bottomrule
\end{tabular}
\end{table}

\subsection{Action Grounding Quality}
\label{sec:eval-b}

We evaluate the grounding pipeline on a curated, researcher-authored probe set of intent descriptions spanning the three behavior bundle pools (378~bundles in total): the talk-bundle pool (INTERACT with dialogue, 258~bundles), the non-talk-bundle pool (INTERACT without dialogue, 90~bundles), and the to-self bundle pool (30~bundles). The probe set is pool-aware: intents were written with reference to the semantic range of the behavior bundle pool, then stratified by difficulty to include (a) close paraphrases of bundle names, (b) indirect or contextual phrasings, (c) semantically adjacent or ambiguous cases, and (d) out-of-scope inputs for testing fallback behavior. We use this probe set to characterize controlled grounding behavior and failure modes of the deployed matcher, not to claim that the sampled intents represent the natural distribution of live player whispers or raw LLM outputs. Each result was evaluated by a human annotator (binary yes/no): is the top-1 matched bundle a reasonable executable interpretation of the intent?

On the talk-bundle pool ($n=45$), the pipeline achieves \textbf{87\%} top-1 accuracy and \textbf{96\%} top-3 accuracy (39/45 and 43/45 respectively; mean cosine similarity~0.64). The six top-1 failures reveal interpretable semantic confusions rather than random noise: some prompts collapse nearby social intents (e.g., ``comfort a friend who seems sad'' retrieves ``Express sadness/disappointment''), while others lose pragmatic directionality or discourse function (e.g., ``check in on how someone is feeling emotionally'' retrieves ``Challenge someone's feelings''). In most of these cases the intended bundle still appears within the top-3, suggesting that embedding retrieval is often close even when the top-ranked match is imperfect.

On the to-self bundle pool ($n=25$), the pipeline achieves \textbf{84\%} top-1 accuracy (21/25), with the same top-3 accuracy because none of the four failures are recovered within the top-3. All four failures are intentionally out-of-scope prompts (e.g., fantastical actions such as flying, teleporting, or breathing fire), so the mixed overall score understates performance on ordinary self-actions. Using the 0.3 threshold as a rough exclusion rule for the most clearly invalid case yields 88\% top-1 accuracy on the remaining 24 probes. In this 100-sample run, one out-of-scope self-action example (``teleport to a different dimension'') falls below the fallback threshold and is routed to the safe default, while the other invalid probes remain above threshold.

On the non-talk-bundle pool ($n=30$), the pipeline achieves \textbf{63\%} top-1 accuracy and \textbf{77\%} top-3 accuracy (19/30 and 23/30 respectively; mean cosine similarity~0.68). The gap relative to the talk-bundle pool is substantial and reflects a qualitatively different error profile: several benign physical-contact or social-gesture intents are pulled toward semantically adjacent intimate bundles. For example, ``give them a warm smile'' retrieves an intimate-gesture bundle; ``pat them gently on the shoulder'' retrieves a neck-touch bundle; ``take a photo with them'' retrieves an explicit-content bundle (Table~\ref{tab:a2}).

This lower accuracy is important but should be interpreted in context. In the deployed interaction pipeline, the talk bundle carries the primary social intent, while the non-talk bundle plays a supplementary role as an accompanying physical gesture and may be omitted altogether in some interactions. As a result, non-talk grounding errors are visible and analytically important, but they are less disruptive to overall social coherence than errors in the talk bundle would be.

To test whether this pattern was primarily caused by explicit-content bundles, we re-ran the same 30 probes with the candidate pool restricted to pglv\,$\leq$\,2 (70 of 90 actions; 20 pglv\,=\,3 bundles removed). Contrary to expectation, grounding did not improve: top-1 accuracy decreased from 63\% to 53\%, and top-3 accuracy from 77\% to 70\% (mean similarity~0.665). Inspection of the seven changed matches reveals why: several pglv\,=\,3 bundles involving physical contact (e.g., body-grab actions) were judged as more reasonable matches than the pglv\,$\leq$\,2 alternatives for physical-force intents such as ``push them away forcefully'' and ``grab their arm to stop them''; removing them replaced these with pglv\,=\,2 intimate-gesture bundles that are less relevant. This result indicates that the difficulty is not specific to the most explicit content level. Rather, the non-talk pool contains a dense embedding neighborhood of body-contact actions spanning all three maturity levels, and semantic interference arises from that structural overlap rather than from any single content tier.

\begin{table}[h]
\scriptsize
\caption{Representative action grounding examples from the 100-item probe set. \textit{Sim} is the top-1 cosine similarity; \checkmark{} / \texttimes{} is human judgment of match quality.}
\label{tab:a2}
\begin{tabular}{@{} >{\RaggedRight}p{3.2cm} >{\RaggedRight}p{3.2cm} l c @{}}
\toprule
Input intent & Matched bundle & Sim & \\
\midrule
ask for their opinion on something & Ask for opinion           & 0.85 & \checkmark \\
suggest a way to improve things & Suggest an improvement       & 0.86 & \checkmark \\
propose a middle-ground compromise & Propose a compromise      & 0.83 & \checkmark \\
express my fear and anxiety about the situation & Express fear\slash anxiety & 0.79 & \checkmark \\
express that I agree with what they said & Express disagreement\slash disapproval & 0.56 & \checkmark \\
comfort a friend who seems sad  & Express sadness\slash disappointment & 0.42 & \texttimes \\
ask them about their future goals and dreams & Teach Knowledge & 0.42 & \checkmark \\
blow a kiss at them             & Run your thumb across their lips & 0.71 & \texttimes \\
\bottomrule
\end{tabular}
\end{table}

Taken together, the results demonstrate grounding feasibility across all three pools. Failure modes are transparent rather than silent: talk-pool mistakes usually remain semantically nearby, and the non-talk accuracy gap reflects a structural representational challenge for physical-contact intents rather than a pool-configuration artifact. The content-level filter (\texttt{MAX\_PGLV}) functions as a runtime safety guardrail; it does not, by itself, resolve grounding ambiguity in the non-talk pool. We emphasize that this section evaluates a controlled grounding benchmark for the deployed matcher, not the ecological distribution of natural player inputs.

\subsection{Whisper Intervention Success}
\label{sec:eval-c}

We evaluated whisper on a controlled, researcher-authored benchmark of 30 cases spanning two use conditions: 20 to-other whispers that asked a character to steer an interpersonal interaction in a particular social direction, and 10 to-self whispers that asked the character to perform a self-directed action. Each case was annotated by a human evaluator as \textit{success}, \textit{partial}, or \textit{failure}. Following the design goal of whisper as \textit{soft intervention}, our primary metric was intervention-aligned rate: whether the resulting action and dialogue preserved the intended steering direction (\textit{success} + \textit{partial}), rather than whether the system obeyed the whisper as a literal command. The goal of this section is therefore to test whether whisper functions as a usable steering signal within bounded autonomy, not to claim that it is universally preferable to direct command interfaces or to passive observation in every gameplay context.

Across all 30 cases, 26 were intervention-aligned (86.7\%), comprising 21 full successes and 5 partial alignments. The strongest results appeared in the to-other condition, where all 20 cases were directionally aligned. Importantly, the 5 non-success to-other cases were all \textit{partial} rather than complete failures: two were labeled \textit{talk\_misalignment}, one \textit{action\_misalignment}, one \textit{over\_softened}, and one \textit{unclear\_whisper}. No to-other case exhibited full \textit{semantic\_drift}. This suggests that when whisper targeted an interpersonal interaction, the system usually preserved the intended social direction even when one side of the realization---the grounded bundle pair or the generated dialogue---remained imperfect or agent-specific. This pattern is consistent with whisper's intended role as guidance rather than direct command execution.

The weaker results appeared in the to-self condition, where 6 of 10 cases were aligned. Here, all 4 failures were labeled \textit{semantic\_drift}. This contrast suggests that the main bottleneck in self-directed whisper was not social-intent interpretation in the language model, but the coverage of the executable to-self bundle pool used by the grounding layer. Both whisper paths are ultimately bounded by the finite behavior bundle pool, but the to-other path has additional compensating flexibility through free-form dialogue generation and two grounded bundle slots, whereas the to-self path depends on a single embedding match into the much smaller to-self bundle pool. Taken together, these results indicate that whisper is effective as a lightweight player-agent steering mechanism within the bounded-autonomy architecture, while also making clear that self-directed whisper remains constrained by the coverage boundary of the to-self bundle pool (Table~\ref{tab:a3_whisper_cases}).

\begin{table}[t]
\footnotesize
\centering
\setlength{\tabcolsep}{3pt}
\caption{Representative outcomes from the controlled whisper benchmark ($N{=}30$).}
\label{tab:a3_whisper_cases}

\begin{tabularx}{\columnwidth}{@{}
    >{\RaggedRight\hsize=0.8\hsize}X
    >{\RaggedRight\hsize=0.9\hsize}X
    >{\RaggedRight\hsize=1.3\hsize}X
    c @{}}
\toprule
\textbf{Whisper} & \textbf{Grounded Action} & \textbf{Talk / Matched Self} & \textbf{Outcome} \\ \midrule

compliment achievement & Praise\slash Compliment + Smile & ``That pitch you closed last week? Brilliant.'' & success \\ \addlinespace[2pt]

ask to share personal & Reveal vulnerabilities + misaligned action & ``Tell me something real about yourself.'' & partial \\ \addlinespace[2pt]

teleport elsewhere & fallback after weak match (Jump) & matched self action: Jump & failure \\ \bottomrule

\end{tabularx}
\end{table}

To verify that whisper was the active steering signal rather than social context alone, we also ran a small cross-whisper check on five to-other cases. For each case, we held the pre-context fixed and substituted a directionally opposing whisper drawn from the same controlled case set along the same social axis (e.g., replacing ``compliment their recent achievement'' with ``push back openly and directly on their perspective''). In all five cases, the generated bundle pair and dialogue shifted to follow the substituted whisper direction. This result does not replace the main benchmark, but it strengthens the interpretation that whisper causally steers the agent's next behavior rather than merely coinciding with context-driven output.

\subsection{Formative Interviews}

To motivate the design of bounded autonomy, we conducted two semi-structured formative interviews (each approximately 45 minutes) with players who had used the system for multiple months. All participants provided informed consent prior to participation.

The interviews showed that players related to these characters as persistent social entities rather than as disposable chat interfaces. P1 described their customized character as a digital ``persona'' and a projection of self into the game world; when autonomous dialogue did not align with what they would say, the character felt ``inauthentic.'' P1 also reported confusion when encountering seemingly random dialogue from other characters, suggesting that social interaction needed to remain legible at the group level rather than only locally fluent. P2 described the character in markedly relational terms---``my child''---and emphasized that ``the personality must remain consistent,'' because changing it made the character feel like ``this ain't my original bibbit.'' The same participant valued the system as a companion ``without forcing daily tasks or payments,'' suggesting a desire for attachment without heavy management overhead. These interviews did not evaluate the proposed mechanisms directly, but they helped ground the paper's problem framing in concrete player concerns about legibility, coherence, and lightweight influence in a shared live social space.

Taken together, these interviews sharpened the design problem that bounded autonomy addresses. Players wanted characters that could participate in a live social world without becoming confusing, incoherent, or difficult to influence when needed. This directly motivated the three mechanisms in Sections~4--6: reply management to keep multi-agent interaction socially coherent, grounding to ensure that generated intent remains executable in-world, and whisper to provide lightweight player steering without collapsing the character into direct command-following.

\section{Limitations and Future Work}

Bounded autonomy suggests that controlling player-owned LLM characters in live multiplayer games is a distinct systems problem, and that relatively simple external mechanisms can be sufficient for deployment when the goal is social stability rather than optimal behavior. The probabilistic decay schedule in Eq.~\ref{eq:decay} is a hand-tuned linear instantiation of the broader principle that deeper reply chains should become progressively less likely to continue, not a claim about an optimal or theoretically unique form; adaptive or learned decay functions that respond to room dynamics---character count, interaction frequency, player activity---could improve performance in more varied settings. The 40-second heartbeat introduces a perception latency that limits reactivity to fast-moving social events, a tradeoff between inference cost and responsiveness. Our current evaluation of agent-agent control likewise emphasizes boundedness rather than arbitration optimality: the decay studies in Section~\ref{sec:eval-a} show that depth-sensitive decay robustly prevents runaway reply cascades under the tested conditions, but they do not establish that relationship-biased reply focus is the best possible routing policy relative to alternatives such as recency-, salience-, or history-based arbitration.

Whisper's effectiveness depends on the underlying LLM following soft guidance reliably. We observe that highly specific or directive whisper inputs produce more consistent results than vague suggestions, suggesting that the interaction technique may benefit from lightweight onboarding that sets player expectations about the ``softness'' of control. More broadly, our current whisper evaluation establishes feasibility of soft steering within the bounded-autonomy pipeline, but it does not directly compare whisper against stronger command interfaces, passive observation, or other intervention designs. Future work could therefore study when players prefer soft steering, how much authorship they attribute to whispered behavior, and whether whisper reduces or redistributes interaction burden over longer play sessions. More generally, richer user studies could contribute beyond evaluating a single mechanism: they could help clarify how players understand bounded autonomy as a whole, what kinds of control they expect from live multiplayer LLM characters, and which tradeoffs among autonomy, coherence, and intervention feel most acceptable in sustained play.

The grounding pipeline uses the pretrained sentence-transformer \texttt{all-mpnet-base-v2}, chosen because it was fast enough for live deployment and shared across grounding and repetition-control paths. This choice improves implementation simplicity and latency, but it also contributes to residual grounding error; a stronger or domain-adapted embedding model would likely improve separation among socially adjacent actions. The grounding evaluation (Section~\ref{sec:eval-b}) makes that boundary concrete: talk-bundle pool grounding reached 87\% top-1 accuracy (39/45), but non-talk-bundle pool grounding reached only 63\% top-1 accuracy (19/30), with several benign physical-contact prompts pulled toward semantically adjacent intimate bundles. A related limitation is that the unrestricted non-talk action pool contains a dense cluster of body-contact actions spanning multiple maturity levels whose embeddings are proximate to many neutral gestures; in our 100-probe evaluation, this pattern appears repeatedly in cases such as smiling, shoulder patting, and taking a photo together. A clear future direction is to treat this pool as a structured retrieval problem rather than a flat nearest-neighbor search: for example, a hierarchical matcher could first separate coarse gesture families before ranking within a narrower subset, and contrastive fine-tuning or hard negative sampling on semantically adjacent body-contact actions could improve discrimination among socially similar but pragmatically distinct bundles. The same boundedness appears in whisper-to-self behavior: because self-directed whispers are grounded by direct embedding match into a finite to-self bundle pool, semantically unsupported inputs can still produce weak nearest-neighbor matches if the fallback threshold is not conservative enough; in the 25-probe self pool, only one clearly invalid example fell below the fallback threshold. The current emotion-exclusion filter is similarly lightweight by design: it removes obviously contradictory bundles using discrete state labels, but it does not model mixed affect, rapid emotional transition, or interpersonal context. Future work could therefore replace binary exclusion with continuous or multi-label affect conditioning so that retrieval can better accommodate ambivalent or socially layered states rather than only hard incompatibilities. A second limitation is methodological: the grounding and whisper probe sets (Sections~\ref{sec:eval-b}--\ref{sec:eval-c}) were manually authored by the researchers rather than sampled from natural LLM outputs or live player whispers in deployed sessions. Because these probe sets were inventory-aware and privacy-preserving, the resulting accuracies are best interpreted as controlled estimates of mechanism feasibility and failure modes, and may be optimistic relative to real-world performance. The whisper evaluation was also annotated by a single evaluator, which was sufficient for this small controlled benchmark but leaves room for future multi-rater validation.

More broadly, bounded autonomy addresses the \textit{control layer} of player-owned LLM characters in live multiplayer games, not the cognition layer. The architecture does not improve a character's understanding of social context, long-term memory, or personality modeling; it constrains and steers whatever behavior the underlying model generates. Richer cognitive architectures could be layered beneath the control mechanisms described here. Our claim is therefore not that bounded autonomy replaces advances in agent cognition, but that controllability, executability, and player steerability become first-order design constraints once player-owned LLM characters operate as gameplay entities in live multiplayer environments.

\section{Conclusion}

We introduced bounded autonomy, a control architecture for making LLM characters workable in live multiplayer games through agent-agent, agent-world, and player-agent control interfaces. Across these interfaces, reply-chain decay, action grounding with fallback, and whisper-based soft steering show how open-ended character behavior can remain executable, steerable, and socially coherent in shared live play without collapsing into either rigid scripting or unconstrained generation. More broadly, this work frames controllability as a distinct runtime control problem for LLM characters in live multiplayer games and provides a concrete exemplar of how that problem can be addressed in practice. We hope bounded autonomy helps establish controllable LLM character play as a productive systems and interaction design space for the HCI and game AI communities.

\bibliographystyle{plainnat}
\bibliography{references}

\end{document}